\documentstyle[prl,aps,preprint]{revtex}

\newcommand{\be}{\begin{equation}}
\newcommand{\ee}{\end{equation}}
\newcommand{\bea}{\begin{eqnarray}}
\newcommand{\eea}{\end{eqnarray}}

\begin{document}
\draft
\widetext

\title{Pursuing Gravitational $S$-Duality}
\author{H. Garc\'{\i}a-Compe\'an$^{a}$\thanks{Present Address:
{\it School of Natural Sciences,
Institute for Advanced Study, Olden Lane, Princeton NJ 08540 USA}. E-mail:
compean@sns.ias.edu}, O. Obreg\'on$^{b}$\thanks{E-mail:
octavio@ifug3.ugto.mx} and C. Ram\'{\i}rez$^c$\thanks{E-mail:
cramirez@fcfm.buap.mx}\\
$^{a}$ {\it Departamento de F\'{\i}sica, Centro de Investigaci\'on y de
Estudios
Avanzados del IPN\\
 P.O. Box 14-740, 07000, M\'exico D.F., M\'exico}\\
$^b$ {\it Instituto de F\'{\i}sica de la Universidad de Guanajuato\\
 P.O. Box E-143, 37150, Le\'on Gto., M\'exico}\\
$^c$ {\it  Facultad de Ciencias F\'{\i}sico Matem\'aticas\\
Universidad Aut\'onoma de Puebla\\
P.O. Box 1364, 72000, Puebla, M\'exico}}

\maketitle

\begin{abstract}
Recently a strong-weak coupling duality in non-abelian non-supersymmetric
theories in four dimensions has been found. An analogous procedure is reviewed, which
allows to find the `dual action' to the gauge theory of dynamical gravity
constructed by the MacDowell-Mansouri model plus the superposition of a
$\Theta$ term.
\end{abstract}


\hspace{1cm}  PACS numbers: 04.20.Cv, 04.50.+h, 11.15.-q, 11.30.Ly

\noindent  Invited paper to appear in the special issue of
the {\it Journal of Chaos, Solitons and Fractals} on ``Superstrings,
M,F,S,...Theory"  (M.S. El Naschie and C. Castro, eds.)

\narrowtext
\newpage

\section{Introduction}

\noindent
{\it S-Duality}

Duality is a notion which in the last years has led to remarkable advances in nonperturbative quantum theory. It is an old known type of symmetry which by interchanging the electric and magnetic fields leaves invariant the vacuum Maxwell equations.
It was extended by Dirac to include sources, with the well known price of
the prediction of monopoles, which appear as the dual particles to the
electrically charged ones and whose existence could not be confirmed up to now. Dirac
obtained that the couplings (charges) of the electrical and magnetical
charged particles are the inverse of each other, $i.e.$ as the electrical
force is `weak', and it can be treated perturbatively, then the
magnetic force among monopoles will be `strong'.

This duality, called `$S$-duality', has inspired  a
great deal of research in the last years. By this means, many
non-perturbative exact results
have been established. In particular, the exact Wilson effective action of
${\cal N}=2$ supersymmetric gauge theories has been computed, showing
the duality symmetries of these effective theories.
It turns out that the dual description is quite adequate to address standard
non-perturbative problems of Yang-Mills theory, such as confinement, chiral
symmetry breaking, etc.

For string theories
as well, quite a lot of new interesting results have been obtained.
Undoubtely, one of the most important and exciting results is the ample evidence
of the existence of a `master' eleven dimensional non-perturbative quantum
theory named $M$-theory. All five ten-dimensional superstring theories,
together with the eleven dimensional supergravity theory, become specific
vacua of the moduli space of $M$-theory. These and other vacua arising
by compactificating  $M$-theory to lower dimensions, are related among
them by a web of dualities (for recent reviews see \cite{sch}).

Roughly speaking, for the effective low energy actions of superstring
theories, strong-weak coupling ($S$-duality) symmetry is realized at the
level of the axion and dilaton moduli. The gravitational sector
appears dynamical with respect to this symmetry. In fact, it is well known
that the inclusion of gravitational corrections is required in order to
test string duality \cite{vwu}, and to check consistency conditions in
$M$-theory \cite{w}.
For the heterotic string theory in ten dimensions,
toroidal compactification to four dimensions on ${\bf T}^6$, gives for
its low energy limit ${\cal N}=4$ super Yang-Mills theory on ${\bf R}^4$.
$S$-duality in the four dimensional theory, arises as a consequence of
superstring duality in six-dimensions between the heterotic theory on
${\bf T}^4$ and the Type IIA on K3 (for a review see \cite{mjd}).
Dynamical gravity arising in string theory can be switched off from the
four-dimensional theory, i.e. gravity enters only as an
non-dynamical external field. From this theory a twisted topological
field theory can be
constructed on a curved four-manifold, which is shown to be $S$-dual
(according to the Montonen-Olive conjecture), by using different formulas
of four-manifolds well known by the topologists \cite{vwd}.

${\cal N}=4$ supersymmetric gauge theories in four dimensions have
vanishing renormalization group $\beta$-function. Montonen and Olive
conjectured that (at the quantum level) these theories would possess an
SL$(2,{\bf Z})$ exact dual symmetry \cite{mo}. Many evidences of this fact
have been found, although a rigorous proof does not exist at present.
For ${\cal
N}=2$ supersymmetric gauge theories in four dimensions, the $\beta$-function
in general does not vanish. So, Montonen-Olive conjecture cannot be
longer valid in the same sense as for ${\cal N}=4$ theories. However, for
theories with SU(2) gauge group, Seiberg and
Witten found that a strong-weak coupling 'effective duality' can be defined on its low energy effective theory for the cases pure
and with matter \cite{sw}. The quantum moduli space of the pure theory is
identificated with a complex plane, the $u$-plane, with singularities
located at the
points $ u=\pm 1,\infty$. It turns out that at $u = \pm 1$ the original
Yang-Mills theory is strongly coupled, but effective duality permits the
weak coupling description at these points in terms of monopoles or dyons
(dual fields). ${\cal N} =1$ gauge theories are also  in the class of
theories with non-vanishing $\beta$-function. More general, for a gauge
group SU$(N_c)$, an effective non-abelian duality is implemented even when
the gauge symmetry is unbroken. It has a non-abelian Coulomb phase.
Seiberg has shown that this non-abelian Coulomb branch is dual to another
non-abelian Coulomb branch of a theory with gauge group SU$( N_f - N_c)$,
where $N_f$ is the number of flavors \cite{s}. ${\cal N}=1$ theories
have a rich phase structure \cite{nati}. Thus, it seems that in supersymmetric
gauge theories strong-weak coupling duality can only be defined for some
particular phases.

For non-supersymmetric gauge theories in four dimensions, the subject of
duality has been explored recently in the abelian as well as in the
non-abelian cases \cite{witt,verlinde,lozano,ganor,mohammedi}. In the
abelian case (on a curved compact
four-manifold $X$)  the $CP$ violating Maxwell theory   partition
 function $Z(\tau)$, transforms as a {\it
modular form} under a finite index subgroup $\Gamma_0(2)$ of SL$(2,{\bf
Z})$ \cite{witt,verlinde}.  The dependence parameter of the partition
function is given by $ \tau = {\theta \over 2 \pi} + {4 \pi i \over e^2}$,
where $e$ is the abelian coupling constant and $\theta$ is the usual theta
angle.  In the case of non-abelian non-supersymmetric gauge theories,
strong-coupling dual theories can be constructed which results in a kind
of dual ``massive" non-linear sigma models
\cite{lozano,ganor,mohammedi,freedman}.  The starting Yang-Mills theory
contains a $CP$-violating $\theta$-term and it turns  out to be equivalent to
the linear combination of the actions corresponding to the self-dual and
anti-self-dual field strengths \cite{witt,verlinde,lozano,ganor,mohammedi}.

\vskip 1truecm

\noindent
{\it Gravitational Duality}

As a matter of fact, string theory constitutes nowadays the only
consistent and phenomenologically acceptable way to quantize gravity.
It contains in its low energy limit Einstein gravity.
Thus, a legitimate question is the one of which is the `dual' theory of
gravity or, more precisely, how gravity behaves under duality
transformations.

It turns out that, at the quantum level, the description of ${\cal N}= 2$
gauge theories on curved manifolds $X$, at the singular points $u=\pm 1$ of the
moduli space, is given by the Maxwell theory coupled to non-dynamical
gravity,

\begin{equation}
 I = I_{M, \theta} + \int_X \bigg( B(\tau) {\rm tr} R \wedge \tilde R +
C(\tau)  {\rm tr} R \wedge R \bigg),
\label{2}
\end{equation}
in such a way that, as shown by Witten \cite{witt}, in order to cancel the
modular anomaly in effective theories, suitable
holomorphic couplings $B(\tau)$ and $C(\tau)$ have to be chosen.

Gravitational analogs of non-perturbative gauge theories were studied
several years ago, particularly in the context of gravitational
Bogomolny bound \cite{gibb}. As recently was shown
\cite{ht},  there are additional non-standard $p$-branes in $D=10$ type II superstring theory
and $D=11$ $M$-theory, and  which are required by U-duality. These
branes were named `gravitational branes' (`G-branes'), because they carry
global charges which correspond to the $ADM$ momentum $P_M$ and to its
`dual', a $(D-5)$-form $K_{M_1 \cdots M_{D-5}},$ which is related to
the NUT charge. These charges are `dual' in the same sense that the
electric and magnetic charges are dual in Maxwell theory, but they
appear in the purely gravitational sector of the theory. Last year, Hull
has shown in \cite{hull} that these global charges $P$ and $K$ arise
 as central charges of the supersymmetric algebra of type II
superstring theory and $M$-theory. Thus the complete spectrum of BPS
states should include a gravitational sector.

An attempt to extend $S$-duality of Yang-Mills theory to gravity was
discussed in \cite{haw}. There, a  clear evidence of identical
quantum properties of electrically and magnetically charged black-holes
was found.

Finally, a different approach to the `gravitational duality' was worked out
in \cite{hh}. In this paper, some techniques of strong-weak coupling
duality for non-supersymmetric Yang-Mills theories were applied to
 the MacDowell-Mansouri dynamical gravity.
These results will be reviewed and discussed here.

\vskip 1truecm

\noindent
{\it Gauge Theories of the Gravitation}

In the past, in the search for the construction of an unified model for
the fundamental interactions, different routes have been followed. A
natural one is to consider higher dimensional models of gravity (and
supergravity)  \cite{review}.  The metric, as well as in Einstein gravity,
is considered the basic field from which the four dimensional
fields that describe the fundamental interactions and gravity itself can be obtained.

In contrast, if one searches for an unified framework of the
non-gravitational fundamental interactions in four space-time dimensions,
the usual way is that these interactions are described in terms
of a connection associated with an internal group leaving space-time
nondynamical.  However, the evolution of the metric of this space-time
should describe just gravity.  Various attempts have been done to
construct Yang-Mills type gauge theories of gravity \cite{west}, where the
{\it basic} fields are the gauge fields of an appropriate group $G$.  The
metric (the tetrad) and the Lorentz connection are obtained only as
components of these fields, so that standard general relativity is a
consequence of the proposed gauge theory.  Following this scheme twenty
years ago, MacDowell and Mansouri (MM) succeeded in constructing a gauge
theory of gravity \cite{mm}.  Pagels gave an Euclidean formulation
\cite{pa}.  Inspired in the MM work, another approach has been constructed
\cite{nos}, based only on a self-dual spin connection, where duality is
defined with respect to the corresponding Lorentz group indices
 and not, as it is usual, with respect to the space-time indices.  This
proposal was generalized to include the MM gauge theory of gravity and the
Pleba\'nski-Ashtekar formulation \cite{ple}.

Knowing the self-dual formulation of the MM theory \cite{mm}, it is
tempting to search if by combining it linearly with the anti-self-dual part,
one gets the MM action plus a $\Theta$-term and then try to proceed as in
Yang-Mills theories to obtain a dual model.  It turns out that all this
works, as it will be shown in this paper.

The paper is organized as follows. Sec. II is devoted to review
non-supersymmetric non-Abelian duality following \cite{ganor,mohammedi}.
In Sec. III we introduce the MM theory and its corresponding theory based on
its self-dual sector \cite{nos}. Then, in Sec. IV,  we describe the
gravitational duality in MM theory. Finally in Sec. V we give our final
remarks and a glimpse of  future research.

\vskip 2truecm

\section{ Strong-Weak Coupling Duality in Non-susy Yang-Mills Theories}

In Yang-Mills theories, the lack of a generalized Poincar\'{e} lemma means
that there is no dual theory in the same sense as for abelian theories
\cite{chan}.  However, for generic Yang-Mills theories, one can follow the
same procedure as in the abelian case \cite{lozano,ganor,mohammedi}.
Usually one constructs a {\it parent Lagrangian} from which one
recuperates the original Lagrangian and its dual, as different limits.

\begin{equation}
L=- \alpha G_{\mu\nu}^a G^{\mu\nu}_ a+G^{\mu\nu}_a
F_{\mu\nu}^a  ,
\label{new}
\end{equation}
$\alpha$ is the coupling constant, $F_{\mu\nu}^a$ is the field
strength given by
$F^a_{ \mu \nu} = \partial_{\mu} A^a_{\nu} -
\partial_{\nu} A^a_{\mu} + f^a_{bc}A^b_{\mu}A^c_{\nu}$,
 and $G^a_{\mu \nu}$ is a Lie algebra-valued Lagrange
multiplier tensor field. If the variables $G$ are integrated out, we get
the usual Yang-Mills Lagrangian $L= {1\over{4 \alpha}}F_{\mu\nu}^a
F_a^{\mu\nu}$.

Further, the euclidean partition function of (\ref{new}), after a partial
integration of the derivatives of $A$, can be rewritten as\footnote{ On
the path integral can be applied the Faddeev-Popov procedure and factor out
the zero modes as an standard quotient of determinants \cite{abe}. The
rest of the path integral is a sum over the classical saddle points.  Here
we will consider only this classical part.}

\begin{equation}
Z=\sqrt{\pi} \int {\cal D}G \, (det M)^{-1/2} \ exp \bigg( - \int (\alpha
G_{\mu\nu}^a G^{\mu\nu}_ a - {M^{-1}}_{\mu\nu}^{ab}\partial_\rho
G^{\nu\rho}_b\partial_\tau G^{\mu\tau }_a )dx \bigg),
\label{z1}
\end{equation}
where $ M^{\mu\nu}_{ab}=f^c_{ab}G^{\mu\nu}_ c$ is the adjoint transformed
of $G$.

This result represents in some sense the dual of the starting Yang-Mills
theory (\ref{new}).  It can be interpreted as a complicated ``massive"
non-linear
sigma model \cite{freedman} with a complicated metric $M$.
 Gannor and Sonnenschein \cite{ganor} showed how to
regain a Yang-Mills theory from this, in such a way that it apparently
leads to the dual theory. Following them, let us define $\bar{A}_\mu^a = -
({M^{-1}})_{\mu\nu}^{ab}\partial_\rho G^{\nu\rho}_b$, from which it turns out that $G$
satisfies the equations of motion $\partial_\nu
G^{\nu\mu}_b+M^{\mu\nu}_{ab} \bar{A}_\nu^a=0$.  In abelian theories, the
Poincar\'e lemma gives the solution to this equation in terms of a vector
potential for the dual of $G$. However, for non-abelian theories, although
$F_{\mu\nu}^a(\bar{A}) = \partial_\mu \bar{A}^a_\nu - \partial_\nu
\bar{A}^a_\mu + f^a_{bc} \bar{A}^b_\mu \bar{A}^c_\nu$ is a solution for
$\tilde G^a_{\mu\nu}$, it is not the most general solution.

Nevertheless, it can be easily seen that the second term in the
exponential of (\ref{z1}) can be rewritten in two different ways, as
$M^{\mu\nu}_{ab}
\bar{A}_\nu^b \bar{A}_\mu^a,$ or as $\partial_\rho G^{\mu\rho}_a
\bar{A}_\mu^a$ plus a total derivative term. Hence, the partition function
can be written as

\begin{equation}
Z=\sqrt{\pi} \int {\cal D}G \, (det M)^{-1/2} \int {\cal D} \bar{A} \
~~exp \bigg( -\int[\alpha G_{\mu\nu}^a G^{\mu\nu}_a - G^{\mu\nu}_a
F_{\mu\nu}^a(\bar{A})]dx \bigg) \delta(2 \bar{A}_\mu^a+2 {M^{-1}}_{\nu\mu}^{ab}
\partial_\rho G^{\rho\nu}_b),
\end{equation}
where the factor 2 in the delta function was introduced for convenience.
If now in this expression the square root of the determinant and the Dirac
delta function are written as exponentials, after a partial integration we
get

\begin{equation}
Z=\sqrt{\pi} \int {\cal D}G {\cal D} \bar{A}{\cal D}\Omega{\cal D}\Lambda
 \,, exp \bigg( \int \{G^{\mu\nu}_a [-\alpha G_{\mu\nu}^a + F_{\mu\nu}^a (\bar{A})-
2{\cal D}_\mu^{(\bar{A})}\Lambda_\nu^a]- {M^{-1}}^{\mu\nu}_{ab}
\Omega_\mu^a\Omega_\nu^b\}dx \bigg) ,
\end{equation}
which after some manipulations turns out to be
\begin{equation}
Z=\pi \int {\cal D}G {\cal D} \tilde{A}
~~exp \bigg( \int G^{\mu\nu}_a [-\alpha G_{\mu\nu}^a + F_{\mu\nu}^a
(\tilde{A})]dx \bigg),
\end{equation}
where $\tilde{A}=\bar{A}-\Lambda.$ This result shows the way back to the
model we started from, as well as the covariance of the partition function
(\ref{z1}) \cite{ganor}.

Further, the procedure stated above can be followed to find (\ref{z1}) for
the non-abelian case with a $CP$-violating $\theta$-term, on the manifold
$X$. The action is given by

\begin{equation}
 I_{YM, \theta} = { 1\over 8 \pi} \int_X d^4 x \sqrt{g} \bigg( { 4\pi
\over g_{YM}^2} {\rm tr} [ F_{\mu\nu}F^{\mu\nu}] + {i \theta \over 4
\pi}\varepsilon_{\mu\nu\rho\sigma} {\rm tr} [F^{\mu\nu} F^{\rho\sigma}] \bigg),
\label{3}
\end{equation}
where $\theta$ is the non-Abelian theta-vacuum and $g_{YM}$ is the
Yang-Mills coupling constant. Equivalently,

\be
 I_{YM, \theta} = { i \over 8 \pi} \int_X d^4 x \sqrt{g}\bigg( \bar{\tau} {\rm
tr} [{^+}F_{\mu\nu} {^+}F^{\mu\nu}] - \tau {\rm tr} [{^-}F_{\mu\nu}
{^-}F^{\mu\nu}]
\bigg),
\label{4}
\ee
here $ \tau = {\theta \over 2 \pi} + { 4 \pi i \over g_{YM}^2}$ and
$\bar{\tau}$ its complex conjugate,  and
${^\pm}F_{mn}$ are the self-dual and anti-self-dual field strengths
respectively and which are given by ${^\pm}F^a_{ \mu \nu} = \partial_{\mu} {^\pm}A^a_{\nu} -
\partial_{\nu} {^\pm}A^a_{\mu} + f^a_{bc}{^\pm}A^b_{\mu}{^\pm}A^c_{\nu}$. Thus, both terms in ( \ref{4}) have the same form as the usual Yang-Mills action and the
$S$-dual Lagrangian to (\ref{3}) can be obtained as the sum of the corresponding dual and anti-dual terms. Thus, one
finds the dual Lagrangian of (\ref{4})  \cite{lozano,ganor,mohammedi} to
be of the form

\be
\begin{array}{ll}
\tilde{I}_{YM,\theta} & = { i \over 8 \pi} \int_X d^4 x \sqrt{g}\bigg( -
{1\over \bar{\tau}} {^+}G^{a}_{\mu\nu} {^+}G^{\mu\nu}_{a} + {1
\over \tau} {^-}G^{a}_{\mu\nu} {^-}G^{\mu\nu}_{a} + 2
({^+}M)^{-1ab}_{\nu\mu} \partial_\rho {^+}G^{\rho \mu}_{a}
\partial_\sigma  {^+}G^{\nu\sigma}_{b} \\
& -2 ({^-}M)^{-1 ab}_{\nu\mu} \partial_\rho
{^-}G^{\rho \mu}_{a} \partial_\sigma {^-}G^{\nu \sigma}_b \bigg),
\label{6}
\end{array}
\ee
where ${^\pm}G$ are, as mentioned, arbitrary two-forms on $X$ and
${^\pm}M$
are, as previously, the adjoints of ${^\pm}G$, correspondingly.

\vskip 2truecm


\section{Gauge Theories of Gravity}

Let us review the MM proposal \cite{mm,pa}.  The starting point in the
construction of this theory is to consider an SO(3,2) gauge theory with a
Lie algebra-valued gauge potential $A^{AB}_\mu$, where the indices $\mu =
0, 1, 2, 3$ are space-time indices and the indices $A, B= 0, 1, 2, 3, 4$
are internal indices.

For the gauge potential $A^{~AB}_\mu$, we may introduce the corresponding
field strength

\begin{equation}
F_{\mu\nu}^{~~AB} = \partial_\mu A_\nu^{~AB} - \partial_\nu A^{AB}_\mu
+ \frac{1}{2} f^{AB}_{CDEF} A^{CD}_\mu A^{EF}_\nu, \label{efe}
\end{equation}
where $f^{AB}_{CDEF}$ are the structure constants of SO(3,2).  The MM approach
chooses in a rather ad hoc way a vanishig ``torsion"

\begin{equation}
F^{a4}_{\mu\nu} \equiv 0, \label{torsion}
\end{equation}
and as a consequence  the gauge group  breaks from SO(3,2)
to SO(3,1)\footnote {Recently Wilczek has argued a most natural way of
avoiding this choice.  Adding consistent potential terms to the MM
Lagrangian, this breaking of the group  can be seen as a spontaneously
symmetry breaking. Thus the Riemann-Einstein structure follows naturally
from the gauge theory and a volume form \cite{wilczek}.}.

Then, the MM action is given by
\cite{mm}

\begin{equation}
I_{MM} = \int_X d^4 x \varepsilon^{\mu\nu\alpha\beta} \varepsilon_{abcd}
F^{ab}_{\mu\nu} F^{cd}_{\alpha\beta},
\label{mma}
\end{equation}
where $a,b,...{\rm etc.}=0,1,2,3,$ $\varepsilon^{\mu \nu \alpha \beta}$
and $\varepsilon_{abcd}$ are the Levi-Civita symbols in four dimensions
defined by $\varepsilon^{0123} = + 1.$

From Eq. ( \ref{efe}), it turns out that

\be
F^{ab}_{\mu \nu} = H^{ab}_{\mu \nu} + K^{ab}_{\mu \nu},
\label{8}
\ee
where

\begin{equation}
H^{ab}_{\mu\nu} = \partial_\mu A^{ab}_{\nu} - \partial_\nu A^{ab}_\mu
+ \frac{1}{2} f^{ab}_{cdef} A^{cd}_{\mu} A^{ef}_{\nu},
\end{equation}
with $f^{ab}_{cdef}$ the structure constants of SO$(3,1)$  and

\begin{equation}
K^{ab}_{\mu \nu} = -  \lambda^2 ( A^{~a4}_\mu
A^{~b4}_\nu - A^{~a4}_\nu A^{~b4}_\mu). \label{k}
\end{equation}

The proposed action of Ref. \cite{nos} is just the (anti)self-dual part of
the MM action (\ref{mma})

\be
 I^{\pm}_{MM} =  \int_X d^4 x \ \varepsilon^{\mu \nu \alpha \beta}
 \varepsilon_{abcd}{^{\pm}}F^{ab}_{\mu \nu}  {^{\pm}}F^{cd}_{\alpha \beta},
\label{mmsd}
\ee
where  ${^{\pm}}A^{ab}_{\mu} = {1 \over 2} (A^{ab}_{\mu} - {i \over 2}
\varepsilon^{ab}_{ \ \ cd}A^{cd}_{\mu})$ are the (anti)self dual parts of
$A^{ab}_{\mu}$ and ${^{\pm}}K^{ab}_{\mu \nu}= {1 \over 2}(K^{ab}_{\mu\nu}
- {1\over 2} i \varepsilon^{ab}~_{cd} K^{cd}_{\mu \nu})$. Then

\be
{^{\pm}}F^{ab}_{\mu \nu} = {^{\pm}}H^{ab}_{\mu \nu} +{^{\pm}}K^{ab}_{\mu \nu},
\label{ammsd}
\ee
where
\begin{equation}
{^{\pm}}H^{ab}_{\mu\nu} = \partial_\mu {^{\pm}}A^{ab}_{\nu} - \partial_\nu
{^{\pm}}A^{ab}_\mu
+ \frac{1}{2} f^{ab}_{cdef} {^{\pm}}A^{cd}_{\mu} {^{\pm}}A^{ef}_{\nu}.
\end{equation}

It was found in \cite{nos} that the action (\ref{mmsd})  leads to a gauge
theory based in the action

\bea
 I^{\pm}_{MM} &=& {1 \over 2}\int_X d^4 x \ \varepsilon^{\mu \nu \alpha \beta}
 \varepsilon_{abcd}
H^{ab}_{\mu \nu} H^{cd}_{\alpha \beta} \pm i \int_X d^4 x \ \varepsilon^{\mu \nu \alpha
\beta} \eta_{ab}
\eta_{cd}H^{ab}_{\mu \nu} H^{cd}_{\alpha \beta}\nonumber\\
&& - 2\lambda^2 \int_X d^4 x \,
\varepsilon^{\mu \nu \alpha \beta} \varepsilon_{abcd} {^{\pm}}K^{ab}_{\mu \nu}
{^{\pm}} H^{cd}_{\alpha \beta} + 2
\lambda^4 \int_X d^4 x \ \varepsilon^{\mu \nu \alpha \beta}
\varepsilon_{abcd} A^{~a4}_\mu
A^{~b4}_\nu A^{~c4}_\alpha A^{~d4}_\beta.
\label{16}
\eea

If now we identify the gauge fields $A_\mu^{ab}$ with the Ricci rotation
coefficients and $A_\mu^{a4}$ with the space-time vierbein, then the first
two terms are of topological nature and are proportional to the Euler
characteristic and the signature of $X$, respectively. The last two terms
represent dynamical gravity with cosmological term in the form of
Pleba\'nski-Ashtekar \cite{ple}.

\vskip 2truecm


\section{Gravitational $S$-duality}

In the section II we have seen that for Yang-Mills theories a kind of
``dual theories" with inverted couplings can be found
\cite{lozano,ganor,mohammedi}.  One starts with the Yang-Mills action plus
a $CP$ violating $\theta$-term.  It is shown that these two terms are
equivalent to a linear combination of the self-dual and anti-self-dual
Yang-Mills actions.

One can search whether the construction of a linear combination of the
corresponding self-dual and anti-self-dual parts of the MM action can be
also reduced to the standard MM action plus a kind of  gravitational
$\Theta$-term
\cite{duff} and, moreover, if by this means one can find the
``dual-theory" associated with the MM theory.

Let us consider the action

\begin{equation}
S=\int_X d^4 x \varepsilon^{\mu\nu\alpha\beta} \varepsilon_{abcd} \bigg(
{^+} \tau {^+}F^{ab}_{\mu\nu} {^+}F^{cd}_{\alpha\beta} - {^-} \tau
{^-}F^{ab}_{\mu\nu} {^-}F^{cd}_{\alpha\beta} \bigg) ,
\label{mmlc}
\end{equation}
where

\begin{equation}
{^\pm}F^{~~ab}_{\mu\nu} =  \frac{1}{2} \bigg( F^{~~ab}_{\mu\nu}
\pm \tilde F^{ad}_{\mu\nu} \bigg),
\label{hodge}
\end{equation}
with $\tilde F^{ab}_{\mu\nu} = - \frac{1}{2} i \varepsilon^{ab}_{~~cd}
F^{cd}_{\mu\nu}$.
It can be shown \cite{sam}, that this action can be rewritten as

\begin{equation}
S=\frac{1}{2} \int d^{4} x \varepsilon^{\mu\nu\alpha\beta} \varepsilon_{abcd}
\bigg[ ({^+}\tau - {^-}\tau) F^{ab}_{\mu\nu} F^{cd}_{\alpha\beta} + (
{^+}\tau + {^-}\tau) F^{ab}_{\mu\nu} \tilde F^{cd}_{\alpha\beta} \bigg].
\label{23}
\end{equation}
MM showed in their original paper \cite{mm} that the first term in this
action reduces to the Euler topological term plus the Einstein-Hilbert
action with a cosmological term. Similarly, as can be seen from
Eq. (\ref{16}), the second term results to be equal to $iP$ where $P$ is
 the Pontrjagin topological term \cite{nos}.
Thus, this term is a genuine $\Theta$ term with $\Theta$ given by the sum
${^+}\tau + {^-}\tau$.

Let us write the action (\ref{mmlc}) as follows
\begin{equation}
S = S_+ + S_-
= \int_X {\cal L}_+ + \int_X{\cal L}_-
\label{seis}
\end{equation}
where ${\cal L}_{\pm}= \pm \varepsilon^{\mu\nu\alpha
\beta}\varepsilon_{abcd} ({^{\pm}}\tau {^\pm}F^{ab}_{\mu\nu}
{^{\pm}}F^{cd}_{\alpha \beta}).$ This additivity property of the
lagrangian is similar to the one which in the Yang-Mills case leads to the
factorization rule of the partition function. However, in this case the
vanishing torsion condition spoils this factorization rule, because it
cannot be separated in self-dual and anti-self-dual parts.

Our second task is to find the ``dual theory", in a similar sense as in
Yang-Mills theories \cite{ganor,mohammedi}.  For that purpose we consider
the parent action

$$ I = \int_X L_+ + \int_X L_-$$
\begin{equation}
= \int_X d^{4}x \varepsilon^{\mu\nu\alpha\beta} \varepsilon_{abcd}
\bigg( c_1{^+} G^{ab}_{\mu\nu} {^+}G^{cd}_{\alpha\beta} + c_3 {^+}
F^{ab}_{\mu\nu} {^+}G^{cd}_{\alpha\beta} \bigg) + \int_X d^{4}x
 \varepsilon^{\mu\nu\alpha\beta} \varepsilon_{abcd}
\bigg( c_2
{^-} G^{ab}_{\mu\nu} {^-} G^{cd}_{\alpha\beta}  + c_4 {^-} F^{ab}_{\mu\nu}
{^-} G^{cd}_{\alpha\beta} \bigg).
\label{parent}
\end{equation}

First of all we will show how to recover the original action (\ref{mmlc})
from the parent action (\ref{parent}). For simplicity we focus in the
partition function for the `self-dual' part of Lagrangian (\ref{parent})
(the anti-self-dual part $L_-$ follows the same procedure).
Integrating out the additional degrees of freedom ${^+}G$, we get the
``effective Lagrangian'' $L^*_+$, {\it i.e.}

\begin{equation}
Z*_+({^+}\tau) = \int {\cal D} {^+}G exp \Bigg( - \int_X L_+
\Bigg)
\label{diez}
\end{equation}
where $Z*_+({^+}\tau)$ is given by
\begin{equation}
Z*_+({^+}\tau) =exp\Bigg( - \int_X L^*_+ \Bigg).
\label{once}
\end{equation}
Performing the Feynman functional integral for the self-dual and
anti-self-dual sectors we get finally

\begin{equation}
L^* = L^*_+ + L^*_-
\label{doce}
\end{equation}

\begin{equation}
L^* =  \varepsilon^{\mu \nu \alpha \beta}\varepsilon_{abcd}
\bigg[ \bigg(- { c_3^2 \over 4 c_1^+} \bigg) {^+}F^{ab}_{\mu \nu}
 {^+}F^{cd}_{\alpha \beta} -
\bigg(- { c_4^2
\over 4 c_1^-} \bigg) {^-}F^{ab}_{\mu \nu} {^-}F^{cd}_{\alpha \beta}\bigg].
\label{trece}
\end{equation}

Which is precisely the  Lagrangian (\ref{mmlc}) from which we
started from, if in (\ref{trece}) we make the identifications

\be
 c_1 = - {1 \over 4 {^+}\tau}, \ \ \ c_2 = - {1 \over 4 {^-}\tau}, \ \ \
c_3 = c_4 =1.
\ee
In order to get the ``dual theory" we follow reference \cite{ganor}.
Therefore, one should start from the partition function

\begin{equation}
\tilde{Z}(\tilde{\tau})= \int {\cal D} {^+}G \, {\cal D} {^-}G \, {\cal D}
A^{ab}_{\mu} \ {\cal D}A^{a4}_{\mu} \
 exp  \bigg( - \int_X L \bigg).
\label{partition}
\end{equation}

Given the definition of ${^\pm}F$ in (\ref{hodge}) and the fact that
${^\pm} G$ are (anti) self-dual fields, one can rewrite the second and
fourth terms in (\ref{parent}) in the following manner

\begin{equation}
4i \varepsilon^{\mu\nu\alpha\beta} F^{ab}_{\mu\nu} \bigg(
c_3 {^+}G_{\alpha\beta ab} - c_4 {^-}G_{\alpha\beta ab} \bigg).
\end{equation}

If we take into account the second term of the expression (\ref{8}) for
$F_{\mu\nu}^{ab}$, and which is quadratic in $A_\mu^{a4}$ (\ref{k}), then
the integral over the components $A^{~a4}_\mu$ is Gaussian,
giving after integration $(det {\bf G})^{-1/2}$, where {\bf G} is a matrix given
by

\begin{equation}
{\bf G}^{\mu\nu}_{ab} = 8 i \lambda^2 \varepsilon^{\mu\nu\alpha\beta}
\bigg( c_3 {^+}G_{\alpha\beta ab} - c_4 {^-}G_{\alpha\beta ab} \bigg).
\end{equation}

Thus, the partition function (\ref{partition}) can be written as

\begin{equation}
Z(\tau)= \int {\cal D}{^+} G\, {\cal D} {^-}G \, {\cal D} A^{ab}_\mu\,\,
(det {\bf G})^{-1/2}\, exp \bigg({- {I\!\!I}} \bigg)
\end{equation}
where

$${I\!\!I} = \int_X {L\!\!L} =$$
\be
=2 i \int_X d^4 x  \varepsilon^{\mu\nu\alpha\beta} \bigg( c_1
{^+}G^{ab}_{\mu\nu} {^+}G_{\alpha\beta ab} - c_2 {^-} G^{ab}_{\mu\nu}
{^-}G_{\alpha\beta ab} + 2 H^{~~ab}_{\mu\nu} ( c_3
{^+} G_{\alpha\beta ab} - c_4 {^-}G_{\alpha\beta ab}) \bigg),
\label{parentsd}
\ee
with $H^{ab}_{\mu\nu} = \partial_\mu A^{ab}_\nu - \partial_\nu A^{ab}_\mu
+ \frac{1}{2} f^{ab}_{cdef} A^{cd}_\mu A^{ef}_\nu$ is the SO(3,1)
field strength.

We now decompose $H^{ab}_{\mu \nu}$ in its self-dual and anti-self-dual
components as follows \begin{equation} H^{ab}_{\mu \nu} = {^+}H^{ab}_{\mu
\nu} +{^-}H^{ab}_{\mu \nu} \end{equation} and integrate out separately in
their (anti)self-dual connections ${^{\pm}}A^{ab}_{\mu}.$

The integration with respect to ${^+}A^{ab}_{\mu}$ (the anti-self-dual
part ${L\!\!L}_-$ follows the same procedure) is given by

\begin{equation}
 exp \Bigg(- \int_X \tilde{{L\!\!L}}_+^* \Bigg) = \int {\cal D} {^+}
 A^{ab}_{\mu} \ exp
\Bigg( - \int_X {L\!\!L}_+ \Bigg).
\label{catorce}
\end{equation}
The part of the Lagrangian (\ref{parentsd}) relevant to this integration
is

\begin{equation}
{L\!\!L}_+ = \dots + 8ic_3 \varepsilon^{\mu \nu \alpha \beta}
{^+}A^{ab}_{\mu} \partial_{\nu}{^+}G_{\alpha \beta ab} + 2i c_3
\varepsilon^{\mu \nu \alpha \beta} c_3 f^{ab}_{cdef} {^+}A^{cd}_{\mu}
{^+}A^{ef}_{\nu} {^+}G_{\alpha \beta ab} + \dots \ \ \ ~~ .
\label{diezseis}
\end{equation}

Hence we finally obtain

\begin{equation}
\tilde{Z}(\tilde{\tau})=\int {\cal D}{^+}G\, {\cal D} {^-}G \,\, (det {\bf
G})^{-1/2} \,\, det ({{^+}M})^{-1/2} det ({{^-}M})^{-1/2} \,\, exp \bigg({-
\int_X\tilde{L\!\!L}}\bigg),
\label{final}
\end{equation}
with

\begin{equation}
\begin{array}{ll}
\tilde{{L\!\!L}} &= \varepsilon^{\mu\nu\alpha \beta} \bigg[ -{1 \over 4
{^+}\tau} {^+}G^{ab}_{\mu \nu} {^+}G_{\alpha \beta ab} + {1 \over 4
{^-}\tau} {^-}G^{ab}_{\mu \nu} {^-}G_{\alpha \beta ab} +
2 \partial_{\nu} {^+}G_{\alpha \beta ab} {({^+}M)}^{-1 \ abcd}_{\mu
\lambda} \epsilon^{\lambda \theta \rho \sigma}
\partial_{\theta} {^+}G_{\alpha \beta cd} \\
&- 2 \partial_{\nu} {^-}G_{\alpha\beta ab} {({^-}M)}^{-1 \ abcd}_{\mu
\lambda} \varepsilon^{\lambda \theta \rho \sigma}
\partial_{\theta} {^-}G_{\rho \sigma cd}\bigg],
\end{array}
\label{tcinco}
\end{equation}
and

\be
{^\pm}M^{\mu\nu cd}_{ab} = \frac{1}{2} \epsilon^{\mu\nu}_{~~\alpha\beta}
\bigg( - \delta^c_a {^\pm}G^{~\alpha\beta d}_{b} + \delta^c_b
{^\pm}G^{~\alpha\beta d}_a + \delta^d_a {^\pm} G^{~\alpha\beta c}_{b}
- \delta^d_b {^\pm} G^{~\alpha\beta c}_a \bigg).
\label{ff}
\ee
The condition of (anti) self-dual factorization for the above Lagrangian is not longer
valid due to the $(det {\bf G})^{-1/2}$ factor in (\ref{final}).

As mentioned, MM were able to reduce their theory to standard general
relativity with cosmological constant plus the Euler topological term by
identifying $A_\mu^{~ab}$ with the spin connection $ \omega^{~ab}_\mu$ and
$A_\mu^{~a4}$ with the tetrad $e^{~a}_\mu$. Obviously then
$H^{~~ab}_{\mu\nu} \equiv R^{ab}_{\mu\nu}$.

The dynamics of the ${^\pm}G$ results rather complicated,  the right
action should de defined taking into account the determinants
appearing in the partition function (\ref{final}).  Actually the
presence of the degrees of freedom of the gauge fields $A_\mu^{~~a4}$ manifest
itself in the condition

\begin{equation}
\varepsilon_{\mu\nu\rho\tau }{\bf G}^{~~\mu\nu\rho}_{\alpha} = 0,
\label{gg}
\end{equation}

It is interesting to remark that the fields $H^{~~ab}_{\mu\nu} \equiv
R^{~~ab}_{\mu\nu}$,   which depend only on $A^{~~ab}_\mu \equiv
\omega^{~~ab}_\mu$, the Ricci rotation coefficients, can also be
considered in an action like (\ref{mmlc}) or (\ref{23}) and then the action
would
be pure topological because the first term in (\ref{23})  would result in
the Euler
characteristic and the second one in the signature.  As has been
shown in \cite{h}, in this last case,  it is possible to define a dual
theory similar
to (\ref{final})   but without  the $(det{\bf G})^{-1/2}$ factor.   Also a kind of
{\it BF} topological gravitational theory  \cite{gary} and its dual
theory have been  analyzed.  This case would correspond to take
$K^{ab}_{\mu\nu}$ in all the MM formalism, in Sec. III, as an  independent
field not having any relation with $A^{~~a4}_\mu$.

\vskip 1truecm

\vskip 2truecm

\section{Final Remarks and Future Research}

In this paper we have defined a gravitational analog of $S$-duality,
following similar procedures to those well known in non-abelian
non-supersymmetric Yang-Mills theories \cite{ganor,mohammedi}.

We have shown how to construct the Lagrangians for `dual' gauge theories
 of gravitation,  in particular for the case of the MM theory. We could
have also studied other approaches as the one of Pagel, based on the O(5) group
\cite{pa}.
We have also argued that the cases of pure topological gravity and
 BF topological gravity \cite{gary} can be understood as a kind of
non-dynamical limits of the case presented here.

We are aware that this analogy was carried out only at the level of the
structure group of the frame bundle over $X$, and not over all the genuine
symmetries which arise in Einstein gravity theories, such as diffeomorphisms
group Diff$(X)$ of $X$.
Related to this, it is interesting to note that in the `dual' Lagrangian
in Eq. (\ref{tcinco}) only partial derivatives of the
$G's$ fields appear, instead of covariant derivatives. However, it can be
shown that following a procedure similar to that presented in the
introduction for Yang-Mills fields, one can get the covariance of the partition functions. As it
has been pointed out by Atiyah \cite{atiyah}, in field theory the $S$-duality
symmetry is a duality between the fundamental homotopy group of the circle
and the space of group characters of representation theory of the circle.
That means a sort of `identification' between algebraic properties of a
topological space.  It would be very interesting to investigate what is
the mathematical interpretation of the `gravitational $S$-duality'.

We feel that in the framework of the already shown ``gravitational
duality" interesting questions arise:

\begin{itemize}

\item The structure that we have shown is as already argued, similar to
that in Yang-Mills theories (see for example \cite{alvarez}).
This points out to the possible existence of ``gravitational monopoles"
and/or solitons.

\item It is of interest to consider supergravity versions of the work
presented here. In general supersymmetry improves the mathematical
consistence, and improves also the duality properties of Yang-Mills
theories \cite{sw,s,nati}.

The self-dual extension of the MM ${\cal N}=1$ supergravity has been
already obtained \cite{nosd}.  So the supersymmetric equivalent to Eq.
(\ref{mmlc}) can be straightforwardly constructed and then its ``$S$-dual"
can be searched for. For ${\cal N} = 2$ (or larger)  supersymmetries we
can try to follow a similar procedure that should allow the construction
of an action with squared \cite{town} self-dual field strengths and one
would expect this action to be a generalization of the Ashtekar
formulation of ${\cal N}=2$ supergravity theories \cite{kunitomo}.

\item It will be also of interest to investigate the relation of the
``gravitational duality" of this paper with the gravitational duality
proposed by Hull by means of the new gravitational branes arising in type
II superstrings and M theory \cite{hull}.

\end{itemize}

The previous points are under current investigation and will be reported
elsewhere.

\vskip 2truecm
\centerline{\bf Acknowledgments}

This work was supported in part by CONACyT grants 3898P-E9608 and
4434-E9406.  One of us (H.G.-C.) would like to thank CONACyT for support
under the program {\it Programa de Posdoctorantes: Estancias Posdoctorales
en el Extranjero para Graduados en Instituciones Nacionales 1997-1998}
and the Institute for Advanced Study for its generous hospitality.

\vskip 2truecm


\end{document}